\documentclass[a4paper,aps,prd,10pt,preprintnumbers,showpacs,twocolumn,superscriptaddress,nofootinbib,amsmath,amssymb]{revtex4-1}
\usepackage{graphicx}
\usepackage{cmap}
\usepackage[utf8]{inputenc}
\usepackage[T1]{fontenc}

\begin{document}

\title{Arbitrarily long-lived quasinormal modes of proper-time flow black holes}
\author{Milena Skvortsova}\email{milenas577@mail.ru}
\affiliation{Peoples’ Friendship University of Russia (RUDN University),
6 Miklukho-Maklaya Street, Moscow, 117198, Russia}

\begin{abstract}

We investigate the quasinormal modes (QNMs) of a massive scalar field in the background of a regular black hole arising from the proper-time flow in asymptotically safe gravity. This quantum-corrected geometry, characterized by a deformation parameter \( q \), smoothly interpolates between a near-extremal regular black hole and the Schwarzschild solution. Employing both the WKB approximation with Padé resummation and time-domain integration, we compute the complex frequencies for various values of the scalar field mass \( \mu \), multipole number \( \ell \), and deformation parameter \( q \). We observe that the real parts of the QNMs increase with the field mass, while the imaginary parts exhibit behavior indicative of long-lived modes. Although quasi-resonances are not detected in the time-domain profiles due to the dominance of late-time tails, we find that the asymptotic decay follows an oscillatory slowly decaying behavior with the power-law envelope. 
\end{abstract}
\pacs{02.60.Lj,04.50.Gh,04.50.Kd,04.70.Bw}
\maketitle


\textbf{Introduction.} The study of quasinormal modes (QNMs) plays a central role in connecting black hole physics with observational gravitational wave astronomy~\cite{LIGOScientific:2016aoc,LIGOScientific:2017vwq}. These characteristic oscillations emerge from the response of a compact object to external disturbances and encode fundamental information about the spacetime geometry. As solutions to linear perturbation equations with specific boundary conditions, QNMs are determined solely by the background metric and the properties of the perturbing field. Because of this, they serve as sensitive probes of both classical and quantum aspects of black hole spacetimes~\cite{Konoplya:2011qq,Kokkotas:1999bd,Berti:2009kk,Bolokhov:2025uxz}.

Incorporating quantum gravity effects into black hole solutions has become a major avenue for exploring physics beyond general relativity. Among the promising approaches is asymptotically safe gravity, which relies on the existence of a non-Gaussian fixed point under renormalization group flow. Recent work in this framework~\cite{Bonanno:2025PTflow} employed the proper-time renormalization group method to construct non-singular black hole geometries. These solutions smoothly interpolate between a regular core and a Schwarzschild-like exterior by encoding scale dependence in Newton’s constant. In particular, they remain free of curvature singularities and exhibit deviations from classical behavior controlled by a dimensionless quantum flow parameter. Recent analyses of axial gravitational perturbations in this setup demonstrated that the quasinormal frequencies can depart significantly from their Schwarzschild counterparts near the critical deformation, with consequences for Hawking evaporation, and photon ring structure \cite{Bonanno:2025PTflow}.

Building on this foundation, it is essential to understand how quantum corrections to the geometry influence not just gravitational dynamics, but also the propagation of more general fields. In \cite{Lutfuoglu:2025blw}, the QNMs and grey-body factors for massless test fields with spins 0, 1, and 1/2 were analyzed in the same PT-flow-modified spacetime. Those studies revealed both similarities and notable differences compared to the gravitational sector, indicating that test field dynamics provide valuable and independent diagnostics of quantum gravitational effects. However, a key limitation of that analysis was the assumption of vanishing field mass, which restricts its applicability to Standard Model particles with negligible rest energy and excludes long-lived resonant states.

In the present work, we take a next step by investigating the behavior of massive scalar fields in the same quantum-corrected black hole background. Massive fields are of particular interest because they can exhibit qualitatively distinct features, such as quasiresonances \cite{Ohashi:2004wr,Konoplya:2004wg,Konoplya:2005hr,Konoplya:2017tvu,Malik:2025ava,Bolokhov:2023ruj,Bolokhov:2023bwm,Skvortsova:2024eqi,Lutfuoglu:2025qkt,Zinhailo:2024jzt} — modes with arbitrarily long lifetimes —and modified late-time tails \cite{Jing:2004zb,Koyama:2001qw,Moderski:2001tk,Rogatko:2007zz,Gibbons:2008rs} due to the presence of an effective mass scale. These effects are especially relevant in contexts where scalar fields play the role of dark matter or arise in extensions of the Standard Model. Moreover, the mass term introduces additional structure in the effective potential, making the resulting QNM spectrum more sensitive to the details of the near-horizon geometry \cite{Konoplya:2024wds}. Initially massless fields acquire effective mass for example in the brane-world models \cite{Seahra:2004fg,Ishihara:2008re} or in the presence of external magnetic field \cite{Konoplya:2007yy,Konoplya:2008hj,Wu:2015fwa,Davlataliev:2024mjl}. After all, the massive fields have potential of observation via the Pulsar Timing Array experiment \cite{NANOGrav:2023gor,NANOGrav:2023hvm} as shown in \cite{Konoplya:2023fmh}. 

We aim to systematically explore how the mass of the scalar field, combined with the strength of the quantum deformation, modifies the quasinormal spectra and late-time wave dynamics. This includes identifying the regimes in which long-lived modes appear, and evaluating the deviations from Schwarzschild results. The interplay between mass and quantum corrections is expected to reveal novel features not captured in the massless case and to offer further insight into the role of effective field content in black hole physics under quantum gravity modifications.

It is worth mentioning that there are a number of other models for quantum corrected black holes whose spectra have been recently extensively investigated in \cite{Konoplya:2024lch,Zhu:2024wic,Konoplya:2022hll,Luo:2024dxl,Konoplya:2025afm,Malik:2024elk,Yang:2024ofe,Stashko:2024wuq,Konoplya:2020fwg,Lutfuoglu:2025hwh,Filho:2024ilq,Konoplya:2023aph,Momennia:2022tug,Lutfuoglu:2025ohb,Bolokhov:2025egl,Konoplya:2017lhs,Dubinsky:2025nxv}.

\vspace{3mm}
\textbf{Black-hole metric and the wave equation.} The search for quantum-improved black hole solutions has long centered on the challenge of resolving classical singularities while retaining viable exterior geometries. One compelling avenue arises from the \emph{asymptotic safety} program~\cite{Weinberg:1979}, which posits that gravitational couplings approach a non-trivial fixed point under renormalization group (RG) flow. This idea, formalized through functional renormalization group techniques~\cite{Reuter:1998, Niedermaier:2006}, has yielded promising evidence for the consistency of gravity as a quantum field theory.

A particularly robust formulation of this flow employs the proper-time regularization scheme~\cite{Bonanno:2020, Bonanno:2024}, which offers both computational control and independence from gauge and parametrization ambiguities. Within this framework, a class of regular black hole spacetimes was recently proposed~\cite{Bonanno:2025PTflow}, in which the Newtonian gravitational coupling runs with the local energy density in a collapsing interior. The interior geometry—governed by this scale-dependent $G(\epsilon)$—is then matched across a hypersurface to an exterior solution that is static, asymptotically flat, and free from curvature singularities.

The exterior spacetime is described by the line element
\begin{equation}
    ds^2 = -f(r)\,dt^2 + \frac{dr^2}{f(r)} + r^2 d\Omega^2,
\end{equation}
with the lapse function given by  \cite{Bonanno:2025PTflow},
\begin{align}\label{eq:metric}
    f(r) &= \frac{3M^2 + q r^4 - M \sqrt{9M^2 + q^2 r^6}}{q r^4} \notag \\
    &\quad + \frac{2}{3} q r^2 \, \mathrm{arctanh}\left( \frac{(q - \sqrt{q^2 + 9M^2/r^6}) r^3}{3M} \right),
\end{align}
where $M$ is the ADM mass, and the parameter $q$ controls the deviation from classical general relativity due to quantum corrections introduced by the renormalization group flow.

This solution provides a smooth interpolation between Schwarzschild geometry and a fully regular core, with quantum corrections becoming negligible in the limit $q \to \infty$. As such, $q$ acts not merely as a deformation parameter, but as a physical scale measuring the strength of UV modifications to the classical theory. The causal structure of the geometry depends intricately on $q$: for sufficiently large $q$, the metric admits two horizons reminiscent of a regular black hole; near the critical threshold \( q_{\text{cr}} \approx 1.37\,M \), the horizons coincide; and for smaller $q$, the object becomes horizonless yet remains nonsingular.

This type of geometry, constructed via matching a physically motivated interior to a non-singular exterior, stands apart from more ad hoc models of regular black holes. It allows for a controlled insertion of quantum gravitational effects with explicit RG flow dynamics, rendering it a tractable and physically grounded framework for phenomenological exploration.

Recent work~\cite{Bonanno:2025PTflow} has shown that axial gravitational perturbations in this background exhibit nontrivial modifications in their quasinormal spectra, especially near the extremal limit. These changes propagate into observable phenomena such as grey-body spectra and Hawking radiation profiles.

To investigate the propagation of a massive scalar field $\Phi$ on the quantum-corrected black hole background, we consider the Klein–Gordon equation
\begin{equation}
    \left( \Box - \mu^2 \right) \Phi = 0,
\end{equation}
where $\mu$ is the mass of the scalar field. Using the standard ansatz
\begin{equation}
    \Phi(t,r,\theta,\phi) = \frac{\psi(r)}{r} Y_{\ell m}(\theta,\phi) e^{-i\omega t},
\end{equation}
and inserting it into the wave equation in the static, spherically symmetric background~\eqref{eq:metric}, one obtains a Schrödinger-like equation for the radial part:
\begin{equation}
    \frac{d^2 \psi}{dr_*^2} + \left[ \omega^2 - V_\text{eff}(r) \right] \psi = 0,
\end{equation}
where $r_*$ is the tortoise coordinate defined by $dr_*/dr = 1/f(r)$. The effective potential $V_\text{eff}(r)$ takes the form
\begin{equation}
    V_\text{eff}(r) = f(r) \left[ \mu^2 + \frac{\ell(\ell+1)}{r^2} + \frac{f'(r)}{r} \right].
\end{equation}
The presence of the mass term introduces long-range modifications to the effective potential, which are expected to qualitatively affect the quasinormal mode spectrum.

\begin{figure*}
\resizebox{\linewidth}{!}{\includegraphics{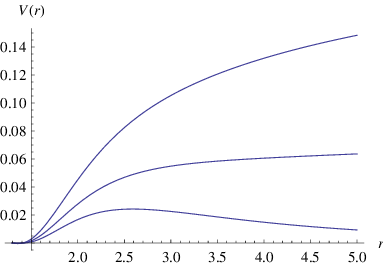}\includegraphics{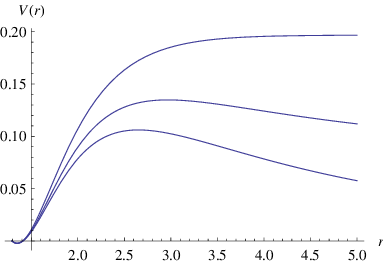}}
\caption{Effective potentials for $\ell=0$ (left) and $\ell=1$ (right) for $\mu=0$ (bottom), $\mu=0.3$, $\mu =0.48$ (top) $M=1$.}\label{fig:Potential}
\end{figure*}


\vspace{3mm}
\textbf{Quasinormal Modes: WKB and time-domain integration methods.} Quasinormal modes (QNMs) characterize the linear response of black holes to external perturbations and are defined as solutions to the corresponding wave equations that satisfy purely ingoing boundary conditions at the event horizon and purely outgoing conditions at spatial infinity. For asymptotically flat black holes, this translates to the following behavior of the radial wave function $\psi(r_*)$ in the tortoise coordinate $r_*$:
\begin{equation}
    \psi(r_*) \sim
    \begin{cases}
        e^{-i\omega r_*}, & r_* \to -\infty \quad (\text{near the horizon}), \\
        e^{+i\omega r_*}, & r_* \to +\infty \quad (\text{spatial infinity}).
    \end{cases}
\end{equation}
These boundary conditions ensure that QNMs represent damped resonances of the system, with complex frequencies $\omega = \omega_R - i \omega_I$, where $\omega_I > 0$ describes the decay rate of the perturbation.

The Wentzel–Kramers–Brillouin (WKB) approximation provides a semi-analytic method to compute quasinormal frequencies for black hole perturbations governed by Schrödinger-like wave equations. In this work, we employ the WKB formula \cite{Schutz:1985km} up to sixth order~\cite{Iyer:1986np,Konoplya:2003ii} with Padé resummation~\cite{Matyjasek:2017psv}, which significantly improves convergence, especially for low-lying modes. The WKB expression reads
\begin{equation}
    \frac{i(\omega^2 - V_0)}{\sqrt{-2V_0''}} - \sum_{j=2}^N \Lambda_j = n + \frac{1}{2},
    \label{eq:WKB}
\end{equation}
where $V_0$ is the peak of the effective potential, $V_0''$ is the second derivative of the potential at the peak with respect to $r_*$, and the $\Lambda_j$ are higher-order correction terms depending on up to $2j$ derivatives of the potential. We use $N=6$ for sixth-order accuracy, and the expression is further improved by applying a Padé approximant to the WKB series, typically of type $(6,6)$ or $(3,3)$, which often stabilizes the result for moderate overtone numbers.

The WKB method is particularly reliable for high multipole numbers $\ell \gg n$, but for fundamental modes or low $\ell$, the results must be compared with more robust time-domain techniques, so that the method was extensively used for finding quasinormal modes 
\cite{Albuquerque:2023lhm,Eniceicu:2019npi,Barrau:2019swg,Konoplya:2005sy,Zinhailo:2019rwd,Chen:2019dip,Skvortsova:2023zca,Konoplya:2006gq,Bronnikov:2019sbx,Dubinsky:2024aeu,Churilova:2021tgn,Konoplya:2001ji,Momennia:2018hsm,Zhidenko:2003wq,Fernando:2012yw,Dubinsky:2025fwv}.

To complement and verify the WKB results, we also employ a fully numerical time-domain integration in the light-cone coordinates $u = t - r_*$ and $v = t + r_*$. The wave equation is discretized using the finite difference scheme originally developed by Price and Pullin~\cite{Gundlach:1993tp}, which is second-order accurate and well-suited for asymptotically flat spacetimes. The discretized update equation reads
\begin{equation}
    \psi(N) = \psi(W) + \psi(E) - \psi(S) - \frac{\Delta^2}{8} V(S) \left[\psi(W) + \psi(E)\right],
\end{equation}
where the grid points $N = (u+\Delta, v+\Delta)$, $S = (u, v)$, $E = (u, v+\Delta)$, and $W = (u+\Delta, v)$ form a diamond-shaped cell in the null plane. The potential $V(S)$ is evaluated at the point $S$.

We initialize the evolution with a Gaussian pulse centered at some finite $r_*$:
\begin{equation}
    \psi(u=0, v) = \exp\left[-\frac{(v - v_c)^2}{2\sigma^2}\right],
\end{equation}
with similar initial values on the adjacent null surface $v = 0$. The resulting time-domain profiles are extracted at fixed $r_*$ and analyzed using the Prony method to identify the dominant quasinormal frequencies from the ringdown signal.
The time-domain integration has been used for finding quasinormal mode, testing the stability of perturbations and determining the asymptotic tails in great number of works \cite{Malik:2024bmp,Skvortsova:2023zmj,Konoplya:2020jgt,Bolokhov:2024ixe,Konoplya:2025uiq,Skvortsova:2024atk,Cuyubamba:2016cug,Churilova:2019qph,Skvortsova:2024wly,Dubinsky:2024hmn,Dubinsky:2025bvf,Qian:2022kaq,Varghese:2011ku,Skvortsova:2024wly,Malik:2025ava,Dubinsky:2024jqi,Konoplya:2013sba,Lutfuoglu:2025bsf,Stuchlik:2025mjj}.

Together, these two complementary methods allow us to reliably compute and cross-check the quasinormal spectra of the massive scalar field in the quantum-corrected background.

\begin{figure*}
\resizebox{\linewidth}{!}{\includegraphics{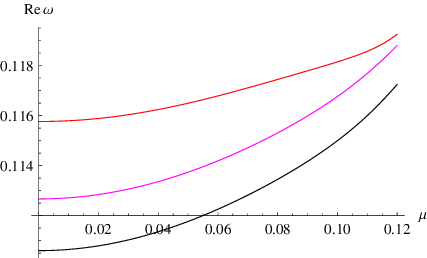}\includegraphics{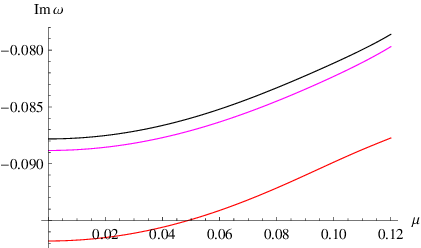}}
\caption{Fundamental quasinormal frequencies as functions of mass $\mu$, $M=1$, $\ell=0$, $q=1.38$ (black), $q=1.6$ (blue), $q=3$ (red).}\label{fig:L0}
\end{figure*}

\begin{figure*}
\resizebox{\linewidth}{!}{\includegraphics{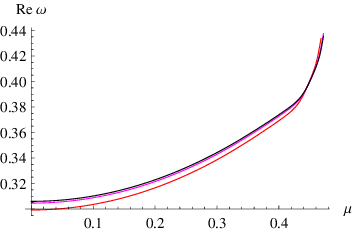}\includegraphics{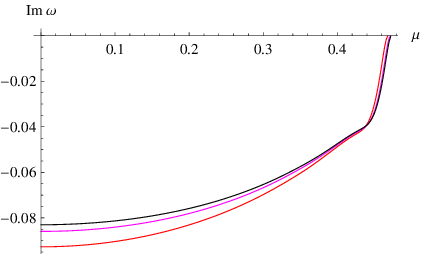}}
\caption{Fundamental quasinormal frequencies as functions of mass $\mu$, $M=1$, $\ell=1$, $q=1.38$ (black), $q=1.6$ (blue), $q=3$ (red).}\label{fig:L1}
\end{figure*}

\begin{figure*}
\resizebox{\linewidth}{!}{\includegraphics{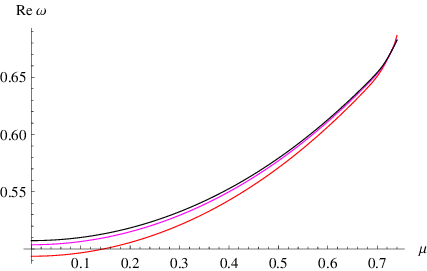}\includegraphics{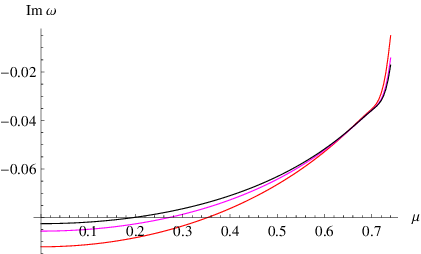}}
\caption{Fundamental quasinormal frequencies as functions of mass $\mu$, $M=1$, $\ell=2$, $q=1.38$ (black), $q=1.6$ (blue), $q=3$ (red).}\label{fig:L2}
\end{figure*}


\begin{figure*}
\resizebox{\linewidth}{!}{\includegraphics{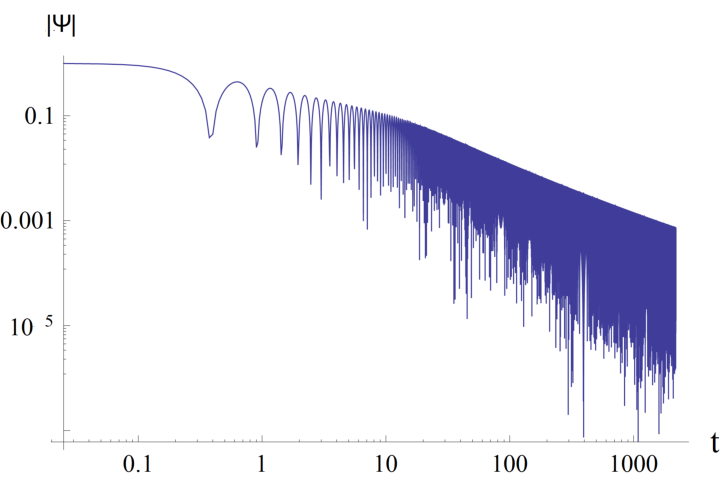}\includegraphics{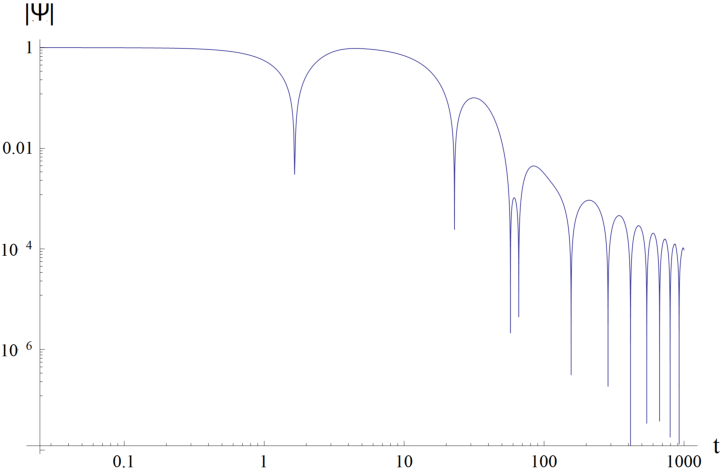}}
\caption{Time domain profile for $q=1.38$, $M=1$, $\ell=0$, $\mu=10$ (left) and $\mu =0.025$ (right).}
\label{fig:TD}
\end{figure*}

\vspace{3mm}
\textbf{Long lived quasinormal modes and late time tails.} The obtained data illustrate the complex interplay between the scalar field mass $\mu$, the quantum deformation parameter $q$, and the multipole number $\ell$ in determining the quasinormal mode (QNM) spectra of the quantum-corrected black hole geometry arising from the proper-time flow.

Figures~\ref{fig:L0}-\ref{fig:L2} show that the real parts of the quasinormal frequencies, $\text{Re}(\omega)$, increase with the field mass $\mu$ across all considered values of the deformation parameter $q$ and multipole numbers $\ell = 0,1,2,3$. This behavior aligns with the general expectation that the introduction of a mass term raises the effective potential barrier, thereby increasing the oscillation frequency. Additionally, stronger quantum deformation (i.e., smaller values of $q$) tends to shift $\text{Re}(\omega)$ upward, especially at intermediate mass scales.

The behavior of the imaginary parts, $\text{Im}(\omega)$, shown in the same figures, exhibits a more intricate structure. For small $\mu$, the damping rate $|\text{Im}(\omega)|$ decreases monotonically as the field mass increases, reflecting the well-known trend toward the quasiresonant regime. At moderate $\mu$, the effective potential becomes broader and lower, leading to partial trapping of the perturbation and thus longer-lived modes. However, as $\mu$ becomes too large, the potential barrier effectively disappears, rendering the WKB method unreliable. In fact, the WKB approximation likely becomes inaccurate already when the damping rate approaches zero. Nevertheless, extrapolation of the curves toward higher $\mu$ suggests the possible existence of arbitrarily long-lived modes in this regime.

The influence of the quantum deformation parameter $q$ on the damping rate is most pronounced near the critical value $q_{\text{cr}} \approx 1.37$, where the two horizons of the quantum-corrected black hole coincide. Near this extremal configuration, the absolute values of the imaginary parts reach their minima. At sufficiently large $\mu$, the frequencies for different values of $q$ converge and become practically indistinguishable within the accuracy of the WKB method.

In contrast, the time-domain profiles do not clearly reveal the quasiresonant behavior, as both the intermediate and asymptotic tails begin to dominate the signal at relatively early times. This limitation arises from the fact that, in asymptotically flat spacetimes, quasinormal modes do not form a complete set and thus do not dictate the evolution of the perturbation at all times.

At asymptotically late times, the decay is no longer governed by quasinormal modes but instead by power-law tails. The observed asymptotic behavior follows the decay law (see fig. \ref{fig:TD})
\begin{equation*}
|\Psi| \sim t^{-5/6} \sin(F(\mu t)), \quad t \rightarrow \infty.
\end{equation*}
The asymptotic decay is appropriate to the Schwarzschild or Reissner-Nordstrom black holes \cite{Koyama:2000hj, Koyama:2001qw}.
At intermediate late times, 
\begin{equation*}
1 \ll t/M \ll (\mu M)^{-3},
\end{equation*}
we observe in fig. \ref{fig:TD} the following behavior:
\begin{equation}
|\Psi| \sim t^{-3/2} \sin(F(\mu t)).
\end{equation}
This corresponds to the well-known intermediate tail regime for massive fields propagating in asymptotically flat Minkowski spacetime.

\vspace{3mm}
\textbf{Conclusions.} We have studied the quasinormal modes (QNMs) of a massive scalar field in the spacetime of a regular black hole solution obtained via the proper-time flow in the framework of asymptotically safe gravity. The resulting geometry, characterized by a quantum deformation parameter \( q \), interpolates between a regular near-extremal black hole and the classical Schwarzschild solution.

Using both the WKB method with Padé resummation and time-domain integration, we have analyzed the behavior of the quasinormal frequencies as functions of the scalar field mass \( \mu \), the deformation parameter \( q \), and the multipole number \( \ell \). We have shown that the real part of the frequencies increases with \( \mu \), while the damping rate decreases,  signaling the emergence of long-lived modes and the breakdown of the WKB approximation near the quasi-resonance regime.

However, the quasi-resonances are not visible in the time-domain profiles, where intermediate and asymptotic tails dominate the signal. We have found that the asymptotic decay at late times follows a power-law tail with an index different from that of the Schwarzschild case, confirming that quantum corrections induce observable modifications in the field evolution. These results provide further insights into the phenomenology of quantum-corrected black holes and highlight the sensitivity of scalar perturbations to both the field mass and the underlying quantum geometry.

It is worth mentioning that the link between quasinormal spectra and grey-body factors, established in \cite{Konoplya:2024lir,Bolokhov:2024otn,Han:2025cal,Malik:2025erb,Lutfuoglu:2025ldc,Malik:2024cgb,Malik:2025dxn}, provides an alternative way to recover scattering characteristics. In particular, the grey-body factors can be directly reconstructed using the numerical quasinormal-mode data obtained in the present analysis. However, these could probably be done with reliable accuracy only in the limit of small mass of the field, when the effective potential has a maximum.

\begin{acknowledgments}
This work was supported by RUDN University research project FSSF-2023-0003.
\end{acknowledgments}

\bibliography{biblio}
\end{document}